\pdfoutput=1
\documentclass[11pt,a4paper]{article}
\usepackage{amssymb}
\usepackage{amsmath}
\usepackage{amsthm}
\usepackage{ragged2e}
\usepackage{amssymb}
\usepackage{siunitx}
\usepackage[left=2.3cm,right=2.3cm,top=2.85cm,bottom=2.55cm]{geometry}

\begin{document}
\title{\Large An Unbounded Fully Homomorphic Encryption Scheme Based On Ideal Lattices Chinese Remainder Theorem}
\author{\hspace{-1cm} Zhiyong Zheng$^{*}$, \small{Engineering Research Center of Ministry of Education for}\\ \hspace{-1.1cm}\small{Financial Computing and Digital Engineering, Renmin University of China,}\\ \hspace{-7.3cm}\small{Henan Academy of Sciences, China}\\
\hspace{-1.4cm} Fengxia Liu$^{\dagger}$, \small{Beijing Advanced Innovation Center for Future Blockchain}\\ \hspace{-2.7cm}\small{and Privacy Computing, Institute of Artificial Intelligence, China,}\\ \hspace{-8.3cm}\small{email: shunliliu@buaa.edu.cn}\\
\hspace{-2cm} Kun Tian$^{\ddagger}$, \small{Engineering Research Center of Ministry of Education for}\\ \small{Financial Computing and Digital Engineering, Renmin University of China, China,}\\ \hspace{-7.8cm}\small{email: tkun19891208@ruc.edu.cn}
}
\date{}
\maketitle

\noindent{\bf{ABSTRACT}}\ \ \  We propose an unbounded fully homomorphic encryption scheme, i.e. a scheme that allows one to compute on encrypted data for any desired functions without needing to decrypt the data or knowing the decryption keys. This is a rational solution to an old problem proposed by Rivest, Adleman, and Dertouzos \cite{32} in 1978, and to some new problems appeared in Peikert \cite{28} as open questions 10 and open questions 11 a few years ago.

\hspace{0.2cm} Our scheme is completely different from the breakthrough work \cite{14,15} of Gentry in 2009. Gentry's bootstrapping technique constructs a fully homomorphic encryption (FHE) scheme from a somewhat homomorphic one that is powerful enough to evaluate its own decryption function. To date, it remains the only known way of obtaining unbounded FHE. Our construction of unbounded FHE scheme is straightforward and noise-free that can handle unbounded homomorphic computation on any refreshed ciphertexts without bootstrapping transformation technique.

\vspace{0.4cm}

\section{Introduction}\label{sec1}
In 1978, Rivest, Adleman, and Dertouzos \cite{32} proposed a concept which has come to be known as fully homomorphic encryption(FHE), at the time they called it a privacy homomorphism. In brief, we write a cryptosystem by $\mathcal{P} \stackrel{f}{\longrightarrow} \mathcal{C} \stackrel{f^{-1}}{\longrightarrow} \mathcal{P}$, where  $\mathcal{P}$ is the plaintexts space, and $\mathcal{C}$ is the ciphertexts space, $f$ is the encryption function depends an public key,  and $f^{-1}$ is the decryption function depends on secret key. Suppose that $\mathcal{P}$ and $\mathcal{C}$ are two commutative rings, and
$$
f^{-1}\left(c_1+c_2\right)=f^{-1}\left(c_1\right)+f^{-1} \left(c_2\right), \ \ f^{-1}\left(c_1 c_2\right)=f^{-1}\left(c_1\right) f^{-1}\left(c_2\right),\ \forall c_1, c_2 \in \mathcal{C},
$$
then $f$ is called a fully homomorphic encryption, or FHE. Under a FHE  $f$, for any polynomials  $p(x) \in \mathcal{C}[x]$, it is easy to see that
$$
f^{-1}\left(p(c)\right)=p^{*}\left(f^{-1}(c)\right), \quad \forall c \in \mathcal{C},
$$
where $p^{*}(x) \in \mathcal{P}[x]$ is the corresponding polynomial  over $\mathcal{P}$.  Since all elementary functions can be approximated by polynomials, thus we can do any desired computation on ciphertexts without the decryption of data.

\subsection{Unbounded FHE Schemes}
Fully homomorphic encryption was known to have abundant applications in cryptography, but for more than three decades no plausibly secure scheme was known. This changed in 2009  when Gentry \cite{14,15}  proposed a candidate FHE scheme based on ideal lattices. The candidate FHE scheme means somewhat homomorphic scheme that supports only a bounded amount of homomorphic  computation on fresh ciphertexts, then, one applies the bootstrapping transformation to convert the scheme into one that can handle unbounded homomorphic computation. Gentry's breakthrough seminal work generated tremendous excitement, and was quickly followed by many works \cite{06,07,08,09,10,11,12,16,17,18,19,33}.
Despite significant advances, bootstrapping is computationally quite expensive, because it involves homomorphically evaluating the entire decryption function. In addition, bootstrapping for unbounded FHE requires one to make a ``circular security" assumption, i.e,  it is secure  to reveal an  encryption of the secret key under itself. So far, such assumptions are poorly understood, and we have little  theoretical evidence to support them, in particular, no worst-case hardness.  Because of this, Peikert \cite{28} put out two open problems relating to unbounded FHE as follows.
\vspace{0.2cm}
\begin{itemize}
\item\ {\bf{Open Question 10:}}  \ Is there an unbounded FHE scheme that does not rely on bootstrapping? Is there a version of bootstrapping that uses a lighter-weight computation than full decryption?

\item\ {\bf{Open Question 11:}} \  Is there an unbounded FHE scheme that can be proved secure solely under a worst-case complexity assumption?
\end{itemize}
\subsection{The FHE Schemes From LWE Distribution }

To date, all known full homomorphic encryption schemes follow the same basic template from Gentry's initial work \cite{14}. In a sequence of work, Brakerski and Vaikuntanathan \cite{08,09} gave a ``second generation" of FHE constructions based on LWE distribution \cite{30,31}. In 2013, Gentry, Sahai, and  Waters \cite{20} proposed another interesting LWE-based FHE scheme that have some unique and advantageous properties. For example, the GSW Scheme can be used to bootstrap with only a small polynomial-factor growth in the error rate. We describe these schemes in details  here.

\vspace{0.3cm}
\begin{itemize}
\item\ \ {\bf{BV FHE Scheme} }\ \ Let $1<t<q$ be two positive integers such that $(t, q)=1$. In the BV system, a secret key $s$ is an LWE secret and encrypt a plaintext $u\in \mathbb{Z}_{t}$ using an LWE sample for modulus $q$. We use the most significant bit encoding to obtain a ciphertext $c$ by
$$
\langle c, s\rangle \equiv _{\chi} \lfloor\frac{q u}{t}\rceil(\text{mod}\  q),
$$
where $\chi$ is a discrete Gauss distribution over $\mathbb{Z}_{q}$, and $\lfloor x\rceil$ is the rounding of real number $x$ to the nearest integer. We use secret key $s$ to decrypt the ciphertext $c$ by
$$
f^{-1}(c)\equiv _{\chi}\lfloor \frac{t}{q} \cdot\langle c, s\rangle\rceil \left(\text{mod}\ t\right).
$$
Obviously, one has $f^{-1}(c)=u$ (see \cite{40}). To explain this scheme is a somewhat FHE scheme, we may use the least significant bit encoding of the message (The equivalence of two encoding is referred  to Appendix A of \cite{04}).
In fact, the ciphertext $c$ satisfies
$$
\langle s, c\rangle\equiv _{\chi} m\ (\text{mod}\ q), \  \text {and }\  m \in\{n t+u \mid n \in \mathbb{Z}\} \cap[-\frac{q}{2}, \frac{q}{2}) .
$$
To decrypt $c$, one just compute $\langle s, c\rangle \in \mathbb{Z}_{q}$, lifts the result to its unique representative $m$ in $\mathbb{Z} \cap [-\frac{q}{2}, \frac{q}{2})$, and the outputs $u\equiv m(\text{mod}\ t)$. It  is easily seen that the homomorphic computation on ciphertext $c$ induces some noises, which, if too large, will destroy the plaintext. Therefore, the bootstrapping technique that re-encrypts a ciphertext and reduces the noise level remains the only known way of building the unbounded FHE schemes.

\end{itemize}

\vspace{0.3cm}
\begin{itemize}
\item\ \ {\bf{GSW FHE Scheme} }\ \ The GSW FHE scheme is presented most simply in terms of the gadget-based trapdoor described
in \cite{01,03,05,24,29}. The heart of GSW scheme are the following additive and multiplicative homomorphisms for tags and trapdoors.  Let
$\bar{A} \in \mathbb{Z}_{q}^{n\times \bar{m}}$ be LWE samples with secret key $\bar{s}\in \mathbb{Z}_{q}^{n-1}$, and for $i=1,2$, let
$$
A_{i}=x_{i}G-\bar{A}R_{i},
$$
where $x_{i}\in \mathbb{Z}_{q} $, $G$ is the block-diagonal gadget matrix of dimension $n\times nl,$ $R_{i} \in \mathbb{Z}_{q}^{\bar{m}\times n}$ are random Gauss matrices over $\mathbb{Z}_{q}$, and $\bar{m}=n+nl$ (see p.50 of \cite{28}). Since  $\left(\begin{array}{l}R_i \\ I_n\end{array}\right)$ is a trapdoor with tag $x_i I_n$ for matrix $\left[\bar{A}, A_i\right] ,$  it is easy to verify that
$$
A_1+A_2=\left(x_1+x_2\right) G-\bar{A}\left(R_1+R_2\right)
$$
and
$$
A_1 G^{-1}\left(A_2\right)=x_1 x_2 G-\bar{A}\underbrace{\left(R_1 G^{-1}\left(A_2\right)+x_1 R_2\right)}_{R}
$$
In other words, $\left(\begin{array}{c}R_1+R_2 \\ I_n\end{array}\right)$ is a trapdoor with tag $\left(x_1+x_2\right) I_n$ for matrix $\left[\bar{A}, A_1+A_2\right]$, and $\left(\begin{array}{l}R \\ I_n\end{array}\right)$ is a trapdoor  with tag $x_1 x_2 I_n$ for matrix $\left[\bar{A}, A_1 G^{-1}\left(A_2\right)\right]$. Even with all of the above techniques,  homomorphic operations always increase the error rate of a ciphertext, by as much as a polynomial factor per operation. Therefore, the schemes described here can only homomorghically evaluate circuits of an a-priori bounded depth.

\end{itemize}

\subsection{The FHE Schemes From Ring-LWE}

Several constructions based on Ring-LWE are today among the most promising FHE candidates. There are three most popular FHE schemes from Ring-LWE: (1) TFHE \cite{Chillotti1,Chillotti2} particularly suitable for combinatorial operations on individual slots and tolerating large noise and thus, large multiplicative depth; (2) B/FV \cite{06,Chen,Fan} allowing to perform large vectorial arithmetic operations as long as the multiplicative depth of the evaluated circuit remains small; (3) HEAAN \cite{Cheon1,Cheon2}---a mixed encryption scheme shown to be very efficient for floating-point computations. We introduce these schemes slightly in details as follows.\\

\begin{itemize}
\item\ \ {\bf{TFHE Scheme}}\ \ TFHE consists of three major encryption/decryption schemes, each represented by a different plaintext space. First, the scheme TLWE encrypts messages over the entire torus $\mathbb{T}$ and produces ciphertexts in $\mathbb{T}^{N+1}$. The other two schemes are:

\hspace*{0.05cm}---\ \ TRGSW encrypts elements of the ring $\mathcal{R}_{\mathbb{Z}}$ (integer polynomials) with bounded $l^{\infty}$-norms (of the corresponding vectors in $\mathbb{Z}^N$ under the natural identification $\mathcal{R}\simeq \mathbb{Z}^N$).

\hspace*{0.05cm}---\ \ TRLWE encrypts elements $\mu$ of the $\mathcal{R}_{\mathbb{Z}}$-module $\mathbb{T}_{\mathcal{R}}$ that can also be viewed as elements of $\mathbb{T}^N$ via the natural bijection $\mathbb{T}_{\mathcal{R}}\simeq \mathbb{T}^{N}$.

\hspace*{0.2cm} There is an external product $\boxdot_{\alpha}$ depending on a noise parameter $\alpha$ (see Corollary 3.14 \cite{Chillotti2}) which yields a FHE module structure on the schemes TRGSW and TRLWE.

\hspace*{0.2cm} In TFHE, TLWE ciphertexts of a message $\mu\in \mathbb{T}$ have the form $(a,b=<s,a>+\mu+e)\in \mathbb{T}^{N+1}$ where $s\in \{0,1\}^N$ is the secret key, $a\in \mathbb{T}^N$ is uniformly random and $e\in \mathbb{T}$ is sampled according to a noise distribution centered at zero. Similarly, for TRLWE, ciphertexts of $\mu\in \mathbb{T}_{\mathcal{R}}$ are of the form $(a,b=s\cdot a+\mu+e)\in \mathbb{T}_{\mathcal{R}}^2$ where $s\in \mathcal{B}$, $a\in \mathbb{T}_{\mathcal{R}}$ is uniformly random and $e\in \mathbb{T}_{\mathcal{R}}$.

\hspace*{0.2cm} The decryption in TLWE (resp. TRLWE) uses a secret $\kappa$-Lipschitz function (here, $\kappa>0$ is small and we mean ``with respect to the $l^{\infty}$-norm on the torus") $\varphi_s: \mathbb{T}^N\times \mathbb{T}\rightarrow \mathbb{T}$ (resp. $\varphi_s: \mathbb{T}_{\mathcal{R}}\times \mathbb{T}_{\mathcal{R}}\rightarrow \mathbb{T}_{\mathcal{R}}$) called phase parametrized by a small (often binary) secret key $s\in \{0,1\}^N$ (resp. $s\in \mathcal{B}$) and defined by $(a,b)\in \mathbb{T}^N\times \mathbb{T}\rightarrow b-<s,a>$ (resp. $(a,b)\rightarrow b-s\cdot a$). The fact that the phase is a $\kappa$-Lipschitz function for small $\kappa \leqslant N+1$ makes the decryption tolerant to errors and allows working with approximated numbers.

\hspace*{0.2cm} Ciphertexts are either fresh (i.e., generated by directly encrypting a plaintext) or they are produced by a sequence of homomorphic operations. In both cases, one views the ciphertext as a random variable depending on the random coins used to generate $a$ and $e$ as well as all random coins used in all these homomorphic operations.

\hspace*{0.2cm} Since $\varphi_s(a,b)=b-s\cdot a=\mu+e$, the decryption $\mu$ and the noise parameter $\alpha$ are the mean and the standard deviation of the phase function $\varphi_s(a,b)$, respectively (here, the mean and standard deviation are computed over the random coins in the encryption algorithm).\\

\item\ \ {\bf{B/FV Scheme}}\ \ In this scheme, the message space is the finite ring $\mathcal{R}_p=\mathbb{Z}_p[x]/<x^N+1>$ for some integer $p$ (typically a power of 2 or a prime number). A message $\mu\in \mathcal{R}_p$ is encrypted on a quotient ring $\mathcal{R}_q$ (for a larger modulus $q$) as a ciphertext $(a,b)\in \mathcal{R}_q^2$ where $a\in \mathcal{R}_q$ is chosen uniformly at random and $b$ is sampled from $\mathcal{D}_{\mathcal{R}_q,\sigma,s\cdot a+\frac{p}{q}\mu}$. Here, $\mathcal{D}_{\mathcal{R}_q,\sigma,\mu}$ is the discrete Gaussian distribution over $\mathcal{R}_q$ centered at $\mu$ with standard deviation $\sigma$ (discrete means that the values are integers only). In addition, $s\in \mathcal{B}$ is the secret key.

\hspace*{0.2cm} Homomorphic addition of two ciphertexts $(a_1,b_1)$ and $(a_2,b_2)$ is achieved by component-wise addition. The idea behind the homomorphic multiplication of two ciphertexts $(a_1,b_1)$ and $(a_2,b_2)$ is a technique referred to as relinearization: one first lifts $(a_i,b_i)\in \mathcal{R}_q^2$ to $(\tilde{a_i},\tilde{b_i})\in \mathcal{R}^2$ where each coefficient is lifted to $[-q/2,q/2)$ and then view of $\mu_i$ as being expressed as a linear polynomial on $s$ (i.e., $\mu_i \sim \frac{p}{q}(b_i+s\cdot a_i)$). One then computes the quadratic polynomial corresponding to the product, namely $\frac{p}{q}(b_1+s\cdot a_1) \cdot \frac{p}{q}(b_2+s\cdot a_2)$, and uses the relinearization described in \cite{Fan} to write this product as $\frac{p}{q}(b+s\cdot a)$ and determine the coefficients $(a,b)\in \mathcal{R}_q$.

\hspace*{0.2cm} The noise amplitude grows by a small factor $O(N)$ on average after each multiplication, so it is a common practice to perform a modulus-rescaling step, that divides and rounds each coefficient as well as the modulus $q$ by the same scalar in order to bring the noise amplitude back to $O(1)$ so that the subsequent operations continue on smaller ciphertexts.\\

\item\ \ {\bf{HEAAN Scheme}}\ \ In this scheme, the message space is the subset of $\mathcal{R}_q$ containing all elements of norm $\leqslant B$ for some bound $B$, where the norm of an element $x\in \mathcal{R}_q$ is defined as $||\tilde{x}||_{\infty}$. Here $\tilde{x}\in \mathcal{R}_{\mathbb{R}}$ is the minimal lift of $x$, i.e., coefficients lifted to $[-q/2,q/2)$. The message is decrypted up to a constant number of least significant bits which are considered as noise.

\hspace*{0.2cm} A HEAAN ciphertext is also a Ring-LWE pair $(a,b)\in \mathcal{R}_q^2$ where $a\in \mathcal{R}_q$ is uniformly random and $b$ is equal to $s\cdot a+\mu$ up to a Gaussian error of small standard deviation. This time, plaintexts and ciphertexts share the same space, so no rescaling factor $p/q$ is used. Multiplication of two messages uses the same formula as in B/FV including relinearization: if both input messages are bounded by $B$ with $O(1)$ noise, the product is a message bounded by $B^2$ with noise $O(B)$, so it is a common practice at this point to perform a modulusrescaling step that divides everything by $B$ to bring the noise back to $O(1)$ (see \cite{Cheon2}). Unlike B/FV, this division in the modulus switching scales not only the ciphertext but also the plaintext by $B$. This can be fixed by adding a (public) tag to the ciphertext to track the number of divisions by $B$ performed.

\end{itemize}

Recently, Boura, Gama, Georgieva, and Jetchev \cite{Boura} proposed a practical hybrid solution for combining and switching between the above three Ring-LWE-based FHE scheme. They achieved it by first mapping the different plaintexts spaces to a common algebra structure and then by applying efficient switching algorithms. This approach has many practical applications, for example, it becomes an integral tool for the recent standardization initiatives of homomorphic schemes and common APIs.

\subsection{Other Related Works}

At Eurocrypt 2010, van Dijk, Gentry, Halevi, and Vaikuntanathan \cite{13} described a fully homomorphic encryption scheme over the integers (see also \cite{10,11,12}). As in Gentry's scheme,  the authors first describe a somewhat homomorphic scheme supporting a limited number of additions and multiplications over encrypted  bits. Then they apply Gentry's squash decryption technique to get a bootstrappable scheme and then Gentry's ciphertext refresh procedure to get a full homomorphic scheme. The main appeal of the scheme (compared to the original Gentry's scheme) is its conceptual simplicity, namely, all operations are done over that integers instead of ideal lattices. However the public key was too large for any practical system. In \cite{11} and \cite{12}, the authors reduced the public key site from $\tilde{\mathcal{O}}\left(\lambda^{10}\right)$ to $\tilde{\mathcal{O}}\left(\lambda^{7}\right)$ by encrypting with a quadratic form in the public key elements, instead of a linear form.

It is the latest schemes without using noise reduction techniques (see \cite{Kogos}). In 2015, Yagisawa proposed a non-associative octonion ring over finite field fully homomorphic encryption scheme \cite{Yagisawa1}. This scheme cryptographic robustness is based on the complexity of multivariate algebraic equations with high degree. In his next paper, he developed some improvements to the scheme \cite{Yagisawa2}. Liu presented a symmetric fully homomorphic encryption scheme based on non-commutative rings \cite{Liu}. Liu's scheme provides arbitrary number of additions and multiplications, and the correct decryption in contempt of the amount of noises reduction mechanisms and based on the approximating-GCD complexity. Noise-free symmetric fully homomorphic encryption scheme was proposed by Li and Wang \cite{Li} that used matrices over non-commutative rings. Needless to say, those  noise-free schemes are not the solutions to problems of FHE scheme, because of using non-associative and non-commutative rings for the plaintexts and ciphertexts.

\subsection{Our contribution}

Our contribution on unbounded  FHE scheme is straightforward and noise-free, which is completely different from Gentry's initial construction.  More precisely, our FHE  scheme can handle unbounded homomorphic computation on any refreshed ciphertexts without bootstrapping transformation. This is an rational solution to open problems on  fully homomorphic encryption.

Our solution comes in three steps. First, we provide a general construction based on ideal lattices and Chinese Remainder Theorem, which includes three efficient algorithms for generating secret key, public key and decryption. The plaintext space is the direct sum of rings $\mathbb{Z}_{t_i}$, where $t_i$ is the one dimensional modulus of each ideal lattice $I_i$ ($1\leqslant i\leqslant m$). The ciphertexts space $\mathbb{Z}^n$ is a commutative ring with the addition and convolutional product $\otimes$ of vectors, needless to say, the algebraic structures for homomorphic computation are simplest. The main ideal is to set up a multiplicative opperation for $\mathbb{Z}^n$, such that $\mathbb{Z}^n$ becomes a commutative ring, therefore, one view $I\subset \mathbb{Z}^n$ both as an ideal and as a lattice in $\mathbb{Z}^n$, in particular, $\mathbb{Z}^n/I$ becomes a quotient ring. Working with this ring $(\mathbb{Z}^n,+,\otimes)$, we can efficiently utilize Chinese Remainder Theorem for generating of public key. Next, we provide an efficient computable practical system based on principal ideal lattices and some basic results from cyclotomic fields \cite{35}. The novelty of this practical system is to establish a connection between a cryptosystem and the ancient hard problem in mathematics: how to find a solution of an algebraic equation with high degree. By the famous Galois theory, there is no general method to find a solution when the degree of an algebraic equation is bigger than 5. Finally, we discuss the secure of our scheme based on the general compact knapsacks problem for arbitrary rings, we show that the security is solely under the worst-case complexity assumption. This is also a solution to Open Question 11 of Peikert \cite{28}.

The generalization of compact knapsack problem for arbitrary ring may be described as follows: given $m$ ring elements $a_{1}, a_{2}, \cdots a_m \in R$ and a target value $b \in R$, find coefficients $x_1, x_2, \cdots, x_m \in X \subset R$ such that $\sum_{i=1}^m a_i x_i=b$. The computational complexity of this problem depends on  the choice of ring $R$ and the size of $X$. This problem is known to be solvable in quasi polynomial time when $R$ is the ring of integers and $X$ is the set of small integers $\left\{0,1, \cdots, 2^{m-1}\right\}$ (see \cite{02} and \cite{22}). Micciancio \cite{22} studied  the knapsack problem when $R$ is an appropriately chosen ring of modulo polynomials and $X$ is the subset of polynomials with small coefficients, he has shown that the complexity is as hard to solve on the average as the worst-case instance of approximating the covering radius of any cyclic lattice within a polynomial  factor.

We generalize this result to any ideal lattices,  such that our FHE scheme is as hard to solve on the average as the worst case of approximating the covering radius of any ideal lattices.

\section{The General Construction}

Let $\mathbb{Z}[x]$ be the ring of integer coefficients polynomials with variable $x,\ \phi(x) \in \mathbb{Z}[x]$, and $\phi(x)=x^n-\phi_{n-1} x^{n-1}-\cdots- \phi_1 x-\phi_0$ with  $ \phi_0 \neq 0$ be a given polynomial, $\langle\phi(x)\rangle$ be the principal ideal generated by $\phi(x)$ in $ \mathbb{Z}[x]$, $ R=\mathbb{Z}[x]/ \langle \phi(x)\rangle$ be the quotient ring. Let $\mathbb{R}^n$ be the $n$ dimensional Euclidean space, and $\mathbb{Z}^n \subset \mathbb{R}^{n}$  be all of integer vectors in $\mathbb{R}^n$. We use column notation for vectors in $ \mathbb{R}^n$.

There is a one to one correspondence between the quotient ring $\mathbb{Z}[x]/ \langle \phi(x)\rangle$ and all the integer vectors $\mathbb{Z}^n$:
\begin{equation*}
a(x)=a_{0} +a_{1}x\cdots a_{n-1}x^{n-1} \in \mathbb{Z}[x]/ \langle \phi(x)\rangle \stackrel{\tau}{\longrightarrow} a=\left(\begin{array}{c}a_{0} \\ a_{1} \\ \vdots \\a_{n-1}\end{array}\right)\in \mathbb{Z}^n, \tag{2.1}
\end{equation*}
we write $\tau(a(x))=a$, or $ \tau^{-1}(a)=a(x)$. Since $\tau\left(a(x)+b(x)\right)=\tau(a(x))+\tau(b(x)), $ $\tau$ is an isomorphism of additive groups in fact. To regard $\tau$ as an isomorphism  of rings, we need to define a multiplicative operator in $\mathbb{Z}^{n}$.  To do this, let the rotation
matrix $H=H_{\phi} $ be given by
$$
H=\left(
\begin{array}{ccc|c}
0 & \cdots & 0 & \phi_0\\
\hline
& & & \phi_1\\
& I_{n-1} & & \vdots \\
& & & \phi_{n-1} \\
\end{array}
\right)
$$
where $I_{n-1}$ is the  unit matrix of $n-1$ dimension.

\vspace{0.3cm}
\textbf{DEFINITION 2.1} \ \ For any $\alpha \in \mathbb{R} ^{n}$, the ideal matrix generated  by $\alpha$ is defined by
\begin{equation*}
H^*(\alpha)=\left[\alpha, H \alpha, \cdots, H^{n-1}\alpha\right] \in \mathbb{R}^{n \times n}.
\end{equation*}

Some basic properties about ideal matrix may be described as the following lemma, its proof  is referred to Lemma $2.5$ of \cite{37}, or Lemma 5.2.4 of \cite{36}.

\vspace{0.3cm}

\textbf{Lemma 2.1} \ \
Let $\alpha=\left(\begin{array}{c}\alpha_0 \\ \alpha_1 \\ \vdots \\ \alpha_{n-1}\end{array}\right)$ and $\beta=\left(\begin{array}{c}\beta_0 \\ \beta_1 \\ \vdots \\ \beta_{n-1}\end{array}\right)$ be any two vectors in $\mathbb{R}^{n}$, then one has

\vspace{0.2cm}
\hspace{2cm} (i)\ \ $H^*(\alpha)=\alpha_0 I_n+\alpha_1 H+\cdots+\alpha_{n-1} H ^{n-1}$;\\

\hspace{2cm} (ii) \ $H^*(\alpha) H^*(\beta)=H^*(\beta) H^*(\alpha)$;\\

\hspace{2cm} (iii) \ $H^*(\alpha) H^*(\beta)=H^*\left(H^*(\alpha) \beta\right)$;\\

\hspace{2cm} (iv) \ $\operatorname{det}\left(H^*(\alpha)\right)=\prod\limits_{i=1}^n \alpha\left(\omega_i\right)$;

\noindent where $\alpha(x)=\tau^{-1}(\alpha)$, and $\omega_1, \omega_2, \cdots,  \omega_n$ are $n$ roots of $\phi(x)$.

\vspace{0.2cm}
By (iv) of Lemma 2.1, $H^{*}(\alpha)$ is an invertible matrix if and only if $\alpha(x)$ and $\phi(x)$ have no common roots in complex numbers field. Now, we may define a multiplicative operator in $\mathbb{Z}^{n}$ in terms of the ideal matrix.

\vspace{0.3cm}
\textbf{DEFINITION 2.2}\ \  Let $\alpha, \beta \in \mathbb{Z}^n$ be two integer vectors, we define the convolutional product $\alpha\otimes \beta$ by
\begin{equation*}
\alpha \otimes \beta=H^*(\alpha) \beta.
\end{equation*}

Obviously, under the convolutional product $\alpha\otimes \beta$, $ \mathbb{Z}^{n}$ becomes a commutative ring with the unit element
$e=\left(\begin{array}{c}1 \\ 0 \\ \vdots \\ 0\end{array}\right) \in \mathbb{Z}^{n}$,
since $\alpha \otimes \beta =\beta \otimes \alpha$ and
\begin{equation*}
\alpha \otimes e=  e \otimes \alpha= \alpha, \ \ \ \forall \alpha \in \mathbb{Z}^{n}.
\end{equation*}
Sometimes, we write this ring by $\left(\mathbb{Z}^{n},+, \otimes\right)=\mathbb{Z}^{n}$.

\vspace{0.3cm}
\textbf{Lemma 2.2} \ \ The correspondence $\tau$ given by (2.1) is a ring isomorphism between $\mathbb{Z}^{n}/\langle\phi(x)\rangle$ and $\mathbb{Z}^{n}$, namely, we have
$$
\mathbb{Z}[x] / \langle\phi(x) \rangle \cong \left(\mathbb{Z}^{n},+, \otimes\right).
$$

\begin{proof}
By Lemma 2.4 of \cite{37} or Lemma 5.2.5 of \cite{36}, it is easy  to see that
$$
\begin{aligned}
& \tau\left(\alpha(x)+\beta(x)\right)=\alpha+\beta=\tau(\alpha(x))+\tau(\beta(x)), \\
& \tau(\alpha(x) \beta(x))=\alpha \otimes \beta=\tau(\alpha(x)) \otimes \tau(\beta(x)) .
\end{aligned}
$$
The conclusion follows immediately.

\end{proof}

According to the definition of ideal lattices \cite{21,22,27,28,38,39}, an ideal lattice $I\subset \mathbb{Z}^n$, just is an ideal of $(\mathbb{Z}^n,+,\otimes)$, we view $I\subset \mathbb{Z}^n$ both as an ideal of $\mathbb{Z}^n$ and as a lattice in $\mathbb{Z}^n$, in particular, $\mathbb{Z}^n/I$ is a quotient ring.

A lattice $I\subset \mathbb{R}^n$ is a discrete additive group of $\mathbb{R}^n$ \cite{26,27,34}, we write $I=\mathcal{L}(B)$ as usual, where $B=[\beta_1,\beta_2,\cdots,\beta_n]\in \mathbb{R}^{n\times n}$ is the generated matrix or a basis of $I$. All lattices we discuss here are the full rank lattice, it means that det$(B)\neq 0$. If $I\subset \mathbb{Z}^n$, then $I$ is called an integer lattice. The Hermite Normal Form base $B$ \cite{23,35} for an integer lattice is an upper triangular matrix $B=(b_{ij})_{n\times n}\in \mathbb{Z}^{n\times n}$ satisfying $b_{ii}\geqslant 1 (1\leqslant i\leqslant n)$ and
\begin{equation*}
b_{ij}=0,\ \text{if}\ 0\leqslant j<i\leqslant n,\ \text{and}\ \ 0\leqslant b_{ij}<b_{ii},\ \text{if}\ 0\leqslant i<j\leqslant n.
\end{equation*}

It is known that there is an unique HNF basis for an integer lattice and its Gram-Schmidt orthogonal basis is a diagonal matrix, more precisely, if $B=[\beta_1,\beta_2,\cdots,\beta_n]$ is the HNF basis of an integer lattice, $B^*=[\beta_1^*,\beta_2^*,\cdots,\beta_n^*]$ is the corresponding orthogonal basis obtained by Gram-Schmidt orthogonal method, then $B^*=\text{diag}\{b_{11},b_{22},\cdots,b_{nn}\}$ is a diagonal matrix (see Lemma 7.26 of \cite{35}).

\vspace{0.3cm}
\textbf{DEFINITION 2.3} \ \ $b_{11}$ is called the one dimensional modulus of an ideal lattice $I=\mathcal{L}(\beta_1,\beta_2,\cdots,\beta_n)$, and denoted by $t(I)=b_{11}$.

\vspace{0.2cm}

Let\  $I\subset (\mathbb{Z}^n,+,\otimes)$ be an ideal lattice, to obtain a set of representative elements for the quotient ring $\mathbb{Z}^n/I$, we use the notation of orthogonal parallelepiped due to Micciancio \cite{23}. The following lemma is referred to Lemma 7.45 and (7.131) of \cite{35}, or \S 4.1 of \cite{23}, and a proof will be appeared in Lemma 2.6 below.

\vspace{0.3cm}
\textbf{Lemma 2.3} \ \ Suppose that $I=\mathcal{L}(B)$ is an full rank ideal of $\mathbb{Z}^n$, $B$ is the HNF basis of $I$, and $B^*=\text{diag}\{b_{11},b_{22},\cdots,b_{nn}\}$ is the corresponding orthogonal basis, then a set of representative elements of the quotient ring $\mathbb{Z}^n/I$ is
\begin{equation*}
\mathcal{F}(I)=\Bigg\{x=\begin{pmatrix} x_1\\x_2\\ \vdots \\ x_n \end{pmatrix}\in \mathbb{Z}^n\ |\ 0\leqslant x_i<b_{ii},\ 1\leqslant i\leqslant n\Bigg\},
\end{equation*}
and $\mathcal{F}(I)$ is called the orthogonal parallelepiped of $I$.

\vspace{0.2cm}

Next, we turn to discuss the ideal lattices and the ring $(\mathbb{Z}^n,+,\otimes)$. Let $\alpha\in \mathbb{Z}^n$ be a given vector, the principal ideal lattice $<\alpha>$ is a principal ideal generated by $\alpha$ in $\mathbb{Z}^n$. It is easily verified that
\begin{equation*}
<\alpha>=\Big\{\alpha \otimes x\ |\ x\in \mathbb{Z}^n\Big\}=\mathcal{L}(H^*(\alpha)).
\end{equation*}
Thus, the generated matrix of a principle ideal lattice $<\alpha>$ just is the ideal matrix generated by $\alpha$.

The operations among the ideal lattices in $(\mathbb{Z}^n,+,\otimes)$ are defined as usual, in particular, the addition and multiplication for two ideals, $I$ and $J$ are defined by
\begin{equation*}
I+J=\Big\{\alpha+\beta\ |\ \alpha\in I,\ \beta\in J \Big\},
\end{equation*}
\begin{equation*}
IJ=\Big\{\sum\limits_{i<\infty}\alpha_i\otimes\beta_i \ |\ \alpha_i\in I,\ \beta_i\in J\Big\},
\end{equation*}
and
\begin{equation*}
I\cap J=\Big\{\gamma\ |\ \gamma\in I, \ \text{and}\ \gamma\in J\Big\}.
\end{equation*}
Since $I+J$, $IJ$ and $I\cap J$ are also the ideals of $\mathbb{Z}^n$, therefore, all of them are ideal lattices.

\vspace{0.3cm}
\textbf{DEFINITION 2.4}\ \  Let $I\subset \mathbb{Z}^n$, $J\subset \mathbb{Z}^n$ be two ideal lattices. If $I+J=\mathbb{Z}^n$, we call $I$ and $J$ to be relatively prime, and denoted by $(I,J)=1$.

\vspace{0.2cm}

It is easy to see that two ideal lattices $I$ and $J$ are relatively prime, if and only if there are $\alpha\in I$ and $\beta\in J$ such that $\alpha+\beta=e$, where $e$ is the unit element of $(\mathbb{Z}^n,+,\otimes)$.

To construct a FHE scheme, we utilize the Chinese Remainder Theorem in the ring $(\mathbb{Z}^n,+,\otimes)$, of which is a well-known theorem in Number Theory.

\vspace{0.3cm}

\textbf{THEOREM 1}(Chinese Remainder Theorem)
\ \ Let $I_1,I_2,\cdots,I_m$ be pairwise relatively prime ideal lattices in $\mathbb{Z}^n$, and $\alpha_1,\alpha_2,\cdots,\alpha_m\in \mathbb{Z}^n$ be $m$ target vectors in $\mathbb{Z}^n$, then there exists a common solution of the following congruences:
\begin{equation*}
x\equiv \alpha_i\ (\text{mod}\ I_i),\ 1\leqslant i\leqslant m,
\end{equation*}
and the solution of $x\in \mathbb{Z}^n$ is unique modulo ideal lattice $I_1\cap I_2 \cdots \cap I_m$.

\vspace{0.2cm}
For arbitrary pairwise relatively prime lattices $I_1,I_2,\cdots,I_m$, it is known that
\begin{equation*}
I_1\cap I_2\cdots \cap I_m=I_1 I_2\cdots I_m.
\end{equation*}
By the above Chinese Remainder Theorem, one has the following consequence immediately.

\vspace{0.3cm}
\textbf{Corollary 2.1} \ \ Suppose that $I_1,I_2,\cdots,I_m$ are pairwise relatively prime ideal lattices in $\mathbb{Z}^n$, then we have the following ring isomorphism
\begin{equation*}
\mathbb{Z}^n/I_1 I_2\cdots I_m \cong \mathbb{Z}^n/I_1 \ \bigoplus \ \mathbb{Z}^n/I_2 \ \bigoplus \ \cdots \ \bigoplus\  \mathbb{Z}^n/I_m. \tag{2.2}
\end{equation*}

\vspace{0.2cm}

The right-hand side of (2.2) is the direct sum of $m$ quotient ring $\mathbb{Z}^n/I_i\ (1\leqslant i\leqslant m)$, and the addition and multiplication of $\mathop{\bigoplus}\limits_{i=1}^m \mathbb{Z}^n/I_i$ are given by
\begin{equation*}
(a_1,a_2,\cdots,a_m)+(b_1,b_2,\cdots,b_m)=(a_1+b_1,a_2+b_2,\cdots,a_m+b_m)
\end{equation*}
and
\begin{equation*}
(a_1,a_2,\cdots,a_m)\otimes(b_1,b_2,\cdots,b_m)=(a_1\otimes b_1,a_2\otimes b_2,\cdots,a_m\otimes b_m).
\end{equation*}

\begin{proof} Since $I_1,I_2,\dots,I_m$ are pairwise prime, by Theorem 1, there are $m$ vectors $A_i$ ($1\leqslant i\leqslant m$) in $\mathbb{Z}^n$ such that
\begin{equation*}
A_i\equiv e\ (\text{mod}\ I_i),\ \text{and}\ A_i\equiv 0\ (\text{mod}\ I_j),\ \text{if}\ j\neq i.
\end{equation*}
For any $\alpha=(\alpha_1,\alpha_2,\dots,\alpha_m)\in \mathbb{Z}^n/I_1 \ \bigoplus \ \mathbb{Z}^n/I_2 \ \bigoplus \ \cdots \ \bigoplus\  \mathbb{Z}^n/I_m$, by Theorem 1 again, there is a unique solution $x\in \mathbb{Z}^n/I_1 I_2\cdots I_m$ such that
\begin{equation*}
\left\{
\begin{array}{c}
x\equiv \alpha_1\ (\text{mod}\ I_1)\\
\vdots\\
x\equiv \alpha_m\ (\text{mod}\ I_m)
\end{array}
\right.
\end{equation*}
We define $f(\alpha)=x$, which is the ring isomorphism from $\mathbb{Z}^n/I_1 \ \bigoplus \ \mathbb{Z}^n/I_2 \ \bigoplus$ $\cdots \ \bigoplus\  \mathbb{Z}^n/I_m$ to $\mathbb{Z}^n/I_1 I_2\cdots I_m$ as we desired. Using $A_1,A_2,\dots,A_m$, one may clearly write down $f$ by
\begin{equation*}
f(\alpha)=f((\alpha_1,\alpha_2,\dots,\alpha_m))=\alpha_1\otimes A_1+\cdots+\alpha_m\otimes A_m.
\end{equation*}
Let $\beta=(\beta_1,\beta_2,\dots,\beta_m)$ be another element of $\mathop{\bigoplus}\limits_{i=1}^m \mathbb{Z}^n/I_i$, it is easy to see that
\begin{equation*}
f(\alpha+\beta)=f((\alpha_1+\beta_1,\alpha_2+\beta_2,\dots,\alpha_m+\beta_m))\
\end{equation*}
\begin{equation*}
\qquad\qquad\qquad\qquad\qquad\ =\alpha_1\otimes A_1+\cdots+\alpha_m\otimes A_m+\beta_1\otimes A_1+\cdots+\beta_m\otimes A_m
\end{equation*}
\begin{equation*}
=f(\alpha)+f(\beta).\qquad\qquad\qquad
\end{equation*}
To verify $f(\alpha\otimes\beta)=f(\alpha)\otimes f(\beta)$, since
\begin{equation*}
f(\alpha)\equiv \alpha_i\ (\text{mod}\ I_i),\ \text{and}\ f(\beta)\equiv \beta_i\ (\text{mod}\ I_i),\ 1\leqslant i\leqslant m.
\end{equation*}
It follows that
\begin{equation*}
f(\alpha)\otimes f(\beta)\equiv \alpha_i \otimes \beta_i\ (\text{mod}\ I_i),\ 1\leqslant i\leqslant m.
\end{equation*}
By the definition of $f$, we have
\begin{align*}
f(\alpha\otimes\beta)&=f((\alpha_1\otimes\beta_1,\alpha_2\otimes\beta_2,\dots,\alpha_m\otimes\beta_m)) \\
&=f(\alpha)\otimes f(\beta).
\end{align*}
This proves that $f$ is a ring isomorphism.
\end{proof}

We extend the inverse isomorphism $f^{-1}$ to the whole space $\mathbb{Z}^n$ by $f^*=f\circ \pi:$
\begin{equation*}
f^*: \mathbb{Z}^n \stackrel{\pi}{\longrightarrow} \mathbb{Z}^n/I_1 I_2\dots I_m \stackrel{f^{-1}}{\longrightarrow} \mathop{\bigoplus}\limits_{i=1}^m \mathbb{Z}^n/I_i,
\end{equation*}
where $\pi$ is the natural homomorphism from $\mathbb{Z}^n$ to its quotient ring $\mathbb{Z}^n/I_1 I_2\dots I_m$. It is easy to see that $f^*$ is a homomorphism from $\mathbb{Z}^n$ to $\mathop{\bigoplus}\limits_{i=1}^m \mathbb{Z}^n/I_i$, and
\begin{equation*}
f^*(f(u))=f^{-1}(f(u))=u,\ \forall u\in \mathop{\bigoplus}\limits_{i=1}^m \mathbb{Z}^n/I_i, \tag{2.3}
\end{equation*}
$f^*$ will play the role of decryption in our scheme, and always write down $f^*$ by $f^{-1}$ in the following discussion.

Since $\mathbb{Z}$ is a ring, we wish to embed this ring into $(\mathbb{Z}^n,+,\otimes)$. To do this, we define an embedding mapping from $\mathbb{Z}$ to $\mathbb{Z}^n$ by
\begin{equation*}
\forall a\in \mathbb{Z}, \ a\rightarrow \overline{a}=\begin{pmatrix} a\\0\\ \vdots\\ 0 \end{pmatrix}\in \mathbb{Z}^n. \tag{2.4}
\end{equation*}

\textbf{Lemma 2.4} \ \ Under the embedding mapping $a\rightarrow \overline{a}$, $\mathbb{Z}$ becomes a subring of $(\mathbb{Z}^n,+,\otimes)$, namely, for any $a\in \mathbb{Z}$, $b\in \mathbb{Z}$, one has $\overline{a+b}=\overline{a}+\overline{b}$ and $\overline{ab}=\overline{a}\otimes \overline{b}$.

\begin{proof}
By (2.4), we have
\begin{equation*}
\overline{a+b}=\begin{pmatrix} a+b\\0\\ \vdots\\ 0 \end{pmatrix}=\begin{pmatrix} a\\0\\ \vdots\\ 0 \end{pmatrix}+\begin{pmatrix} b\\0\\ \vdots\\ 0 \end{pmatrix}=\overline{a}+\overline{b}
\end{equation*}
and
\begin{equation*}
\overline{a}\otimes\overline{b}=\begin{pmatrix} ab\\0\\ \vdots\\ 0 \end{pmatrix}=\overline{a\cdot b},
\end{equation*}
the lemma follows at once.
\end{proof}

\textbf{Lemma 2.5}\ \ Let $I\subset \mathbb{Z}^n$ be an ideal lattice, and $t=t(I)$ be its one dimensional modulus given by DEFINATION 2.3. Then, for any $a,b\in \mathbb{Z}$, we have
\begin{equation*}
a\equiv b\ (\text{mod}\ t)\Longleftrightarrow\overline{a}\equiv \overline{b}\ (\text{mod}\ I).
\end{equation*}

\begin{proof}
By assumption, $I$ is a full rank lattice and its HNF basis may be written by
\begin{equation*}
B=\begin{pmatrix}
b_{11} & * & * & *\\
& b_{22} & * & *\\
& 0 & \ddots & *\\
 & & & b_{nn}
\end{pmatrix}.
\end{equation*}
If $\overline{a}\equiv \overline{b}\ (\text{mod}\ I)$, then there is a vector $x=\begin{pmatrix} x_1\\ \vdots \\x_n \end{pmatrix}\in \mathbb{Z}^n$ such that
\begin{equation*}
\overline{a}-\overline{b}=Bx=\begin{pmatrix} a-b\\ 0 \\ \vdots \\ 0 \end{pmatrix}.
\end{equation*}
Since $b_{nn} x_n=0$, and $b_{nn}\geqslant 1$, we have $x_n=0$, similarly, since $b_{n-1,n-1}x_{n-1}+b_{n-1,n}x_n=0$, it follows that $x_{n-1}=0$. Therefore, we have $x_n=x_{n-1}=\cdots=x_2=0$, and $a-b=x_1 b_{11}=x_1 t\Rightarrow a\equiv b\ (\text{mod}\ t)$. Conversely, if $a\equiv b\ (\text{mod}\ t)$, then $a-b=qt$, and
\begin{equation*}
\overline{a}-\overline{b}=B\begin{pmatrix} q\\ 0 \\ \vdots \\ 0 \end{pmatrix}\in I\Longrightarrow \overline{a}\equiv \overline{b}\ (\text{mod}\ I).
\end{equation*}
The assertion is true.
\end{proof}

For arbitrary given pairwise relatively prime ideal lattices $I_1,I_2,\cdots,I_m$, one uses the Chinese Remainder Theorem to generate the public key $A_1,A_2,\cdots,A_m$, where $A_i\in \mathbb{Z}^n\ (1\leqslant i\leqslant m)$ such that
\begin{equation*}
A_i\equiv e\ (\text{mod}\ I_i),\  \text{and}\ A_i\equiv 0\ (\text{mod}\ I_j), \ \text{if}\ j\neq i, \tag{2.5}
\end{equation*}
where $e=\begin{pmatrix} 1\\ 0 \\ \vdots \\ 0 \end{pmatrix}\in \mathbb{Z}^n$ is the unit element of $(\mathbb{Z}^n,+,\otimes)$.

Now, we describe a scheme for unbounded fully homomorphic encryption as follows.

\begin{table}[h]
\begin{tabular}{l}
\hline
{{\bf{\hspace*{2cm} Algorithm 1: The General Unbounded FHE Scheme}}}\\
\hline
\\
$\bullet$\ \ {\bf{Secret Key:}}\ \ One selects $m$ pairwise relatively prime ideal lattices $I_1,I_2,\cdots,I_m$ in \\
\ \ \ \ \ \ \ \ \ \ \ \ \ \ \ \ \ \ \ \ \ \ \  $\mathbb{Z}^n$ as the decryption key. \\

$\bullet$\ \ {\bf{Public Key:}}\ \  Let $t_i=t(I_i)$ be the one dimensional modulus of $I_i\ (1\leqslant i\leqslant m)$. The\\
\ \ \ \ \ \ \ \ \ \ \ \ \ \ \ \ \ \ \ \ \ \ \ plaintexts space is the direct sum of ring $\mathcal{P}=\mathop{\bigoplus}\limits_{i=1}^m \mathbb{Z}_{t_i}$, the addition and\\
\ \ \ \ \ \ \ \ \ \ \ \ \ \ \ \ \ \ \ \ \ \ \ multiplication in $\mathcal{P}$ are given by
\\
\vspace{0.05cm} $\hspace{2.8cm} (a_1,a_2,\cdots,a_m)+(b_1,b_2,\cdots,b_m)=(a_1+b_1,a_2+b_2,\cdots,a_m+b_m),$
\\
$\hspace{3.5cm} (a_1,a_2,\cdots,a_m)\cdot(b_1,b_2,\cdots,b_m)=(a_1 b_1,a_2 b_2,\cdots,a_m b_m).$
\\
\ \ \ \ \ \ \ \ \ \ \ \ \ \ \ \ \ \ \ \ \ \ \ The public key for encryption is $\{A_1,A_2,\cdots,A_m\}\subset \mathbb{Z}^n$, and each $A_i$\\
\ \ \ \ \ \ \ \ \ \ \ \ \ \ \ \ \ \ \ \ \ \ \ is given by (2.5). \\

$\bullet$\ \ {\bf{Encryption:}}\ \ For any plaintext $u=(u_1,u_2,\cdots,u_m)\in \mathcal{P}=\mathop{\bigoplus}\limits_{i=1}^m \mathbb{Z}_{t_i}$, the encryption\\
\ \ \ \ \ \ \ \ \ \ \ \ \ \ \ \ \ \ \ \ \ \ \ function $f$ is given by
\\
$\hspace{4.2cm} c=f(u)=\overline{u_1}\otimes A_1+\overline{u_2}\otimes A_2+\cdots+\overline{u_m}\otimes A_m,\qquad\quad\ (2.6)$
\\
\ \ \ \ \ \ \ \ \ \ \ \ \ \ \ \ \ \ \ \ \ \ \ where $\overline{u_i}$ is the embedding of $u_i$. \\

$\bullet$\ \ {\bf{Decryption:}}\ \
For any ciphertext $c\in \mathbb{Z}^n$, we use the secret key $I_1,I_2,\cdots,I_m$ to decrypt\\
\ \ \ \ \ \ \ \ \ \ \ \ \ \ \ \ \ \ \ \ \ \ \ $c$. Since for every $i$, $1\leqslant i\leqslant m$, we have $c\equiv \overline{u_i}\ (\text{mod}\ I_i)$, and $c$ mod $I_i$ is\\
\ \ \ \ \ \ \ \ \ \ \ \ \ \ \ \ \ \ \ \ \ \ \ a unique vector in the orthogonal parallelepiped $\mathcal{F}(I_i)$ of $I_i$, thus one has\\
\ \ \ \ \ \ \ \ \ \ \ \ \ \ \ \ \ \ \ \ \ \ \ $c$ mod $I_i=\overline{u_i}$, and by (2.3), we have
\\
$\hspace{5.3cm} f^{-1}(c)=(u_1,u_2,\cdots,u_m)=u.$
\\ \\
\hline
\end{tabular}
\end{table}

To verify the homomorphic addition and multiplication for the above scheme, by Lemma 2.4, $\mathbb{Z}$ is a subring of $(\mathbb{Z}^n,+,\otimes)$. By Lemma 2.5, we can embed the direct sum $\mathop{\bigoplus}\limits_{i=1}^m \mathbb{Z}_{t_i}$ into $\mathop{\bigoplus}\limits_{i=1}^m \mathbb{Z}^n/I_i$, so that $\mathop{\bigoplus}\limits_{i=1}^m \mathbb{Z}_{t_i}$ also is a subring of $\mathop{\bigoplus}\limits_{i=1}^m \mathbb{Z}^n/I_i$. Therefore, the property of fully homomorphism directly follows by $f^*$ (or $f^{-1}$) a ring homomorphism:
\begin{equation*}
\mathop{\bigoplus}\limits_{i=1}^m \mathbb{Z}_{t_i}\hookrightarrow \mathop{\bigoplus}\limits_{i=1}^m \mathbb{Z}^n/I_i \stackrel{f}{\longrightarrow}   \mathbb{Z}^n/I_1 I_2 \dots I_m \stackrel{f^*}{\longleftarrow} \mathbb{Z}^n.
\end{equation*}
More precisely, let $c_1=f(u)$, and $c_2=f(v)$ be arbitrary two ciphertexts, where $u=(u_1,u_2,\cdots,u_m)\in \mathop{\bigoplus}\limits_{i=1}^m \mathbb{Z}_{t_i}$, and $v=(v_1,v_2,\cdots,v_m)\in \mathop{\bigoplus}\limits_{i=1}^m \mathbb{Z}_{t_i}$, we note that for all $i$, $1\leqslant i\leqslant m$,
\begin{equation*}
c_1+c_2\equiv \overline{u_i}+\overline{v_i}\ (\text{mod}\ I_i),\ \text{and}\ c_1\otimes c_2 \equiv \overline{u_i}\otimes \overline{v_i}\ (\text{mod}\ I_i).
\end{equation*}
By Lemma 2.4, it follows that
\begin{equation*}
c_1+c_2\equiv \overline{u_i+v_i}\ (\text{mod}\ I_i),\ \text{and}\ c_1\otimes c_2 \equiv \overline{u_i v_i}\ (\text{mod}\ I_i).
\end{equation*}
Therefore, by Lemma 2.5, we have
\begin{equation*}
f^{-1}(c_1+c_2)=(u_1+v_1,u_2+v_2,\cdots,u_m+v_m)
\end{equation*}
\begin{equation*}
\qquad\qquad\qquad\ \ =(u_1,u_2,\cdots,u_m)+(v_1,v_2,\cdots,v_m)
\end{equation*}
\begin{equation*}
\ =f^{-1}(c_1)+f^{-1}(c_2),
\end{equation*}
and
\begin{equation*}
f^{-1}(c_1\otimes c_2)=(u_1v_1,u_2v_2,\cdots,u_mv_m)\qquad\qquad
\end{equation*}
\begin{equation*}
\qquad\qquad\quad=(u_1,u_2,\cdots,u_m)(v_1,v_2,\cdots,v_m)
\end{equation*}
\begin{equation*}
=f^{-1}(c_1)\cdot f^{-1}(c_2).\ \
\end{equation*}
This is an unbounded fully homomorphic encryption as we desirable.

How to decrypt the ciphertext $c\in \mathbb{Z}^n$ using the secret key $I_1,I_2,\dots,I_m$ in our algorithm for the general unbounded FHE scheme? Here we give a lemma to show that there is only one vector $\overline{u_i}\in \mathbb{Z}^n$ in the orthogonal parallelepiped $\mathcal{F}(I_i)$ of $I_i$ such that $c\equiv \overline{u_i}\ (\text{mod}\ I_i)$, and give an algorithm to calculate $\overline{u_i}$ in detail.

\vspace{0.2cm}
\textbf{Lemma 2.6}\ \ Given an ideal lattice $I$, for any vector $c\in \mathbb{Z}^n$, there is only one vector $w\in \mathcal{F}(I)$ such that $c\equiv w\ (\text{mod}\ I)$.

\begin{proof} Assume $B=[\beta_1,\beta_2,\dots,\beta_n]$ is the HNF basis of $I$, and $B^*=[\beta_1^*,\beta_2^*,\dots,\beta_n^*]$ is the corresponding orthogonal basis of $B$. Write $c$ as the linear combination of $[\beta_1^*,\beta_2^*,\dots,\beta_n^*]$, i.e.
\begin{equation*}
c=\sum\limits_{i=1}^n c_i \beta_i^*,\ c_i=\frac{<c,\beta_i^*>}{<\beta_i^*,\beta_i^*>}.
\end{equation*}
Let $w=\sum\limits_{i=1}^n c_i \beta_i^*-\sum\limits_{i=1}^n k_i \beta_i$, $k_1,k_2,\dots,k_n\in \mathbb{Z}$, then $c-w=\sum\limits_{i=1}^n k_i \beta_i\in I$. Next, we prove that there is only one group of integers $k_1,k_2,\dots,k_n$ such that $w\in \mathcal{F}(I)$, i.e.
\begin{equation*}
w_i=\frac{<w,\beta_i^*>}{<\beta_i^*,\beta_i^*>}\in [0,1),\ \forall 1\leqslant i\leqslant n.
\end{equation*}
Firstly, we determine the value of $k_n$. Since
\begin{equation*}
w_n=\frac{<w,\beta_n^*>}{<\beta_n^*,\beta_n^*>}=c_n-k_n\frac{<\beta_n,\beta_n^*>}{<\beta_n^*,\beta_n^*>}=c_n-k_n,
\end{equation*}
therefore, when $k_n=[c_n]$, here $[x]$ means the largest integer no more than real number $x$, we have $w_n\in [0,1)$. Secondly, we determine $k_{n-1}$. Note that
\begin{equation*}
w_{n-1}=\frac{<w,\beta_{n-1}^*>}{<\beta_{n-1}^*,\beta_{n-1}^*>}=c_{n-1}-k_{n-1}-k_n\frac{<\beta_n,\beta_{n-1}^*>}{<\beta_{n-1}^*,\beta_{n-1}^*>},
\end{equation*}
so there is only one integer
\begin{equation*}
k_{n-1}=\left[c_{n-1}-[c_n]\frac{<\beta_n,\beta_{n-1}^*>}{<\beta_{n-1}^*,\beta_{n-1}^*>}\right]
\end{equation*}
such that $w_{n-1}\in [0,1)$. Similarly, we could determine any $k_i$, $\forall 1\leqslant i\leqslant n-1$,
\begin{equation*}
w_i=\frac{<w,\beta_i^*>}{<\beta_i^*,\beta_i^*>}=c_i-k_i-\frac{<\sum\limits_{j=i+1}^n k_j \beta_j,\beta_i^*>}{<\beta_i^*,\beta_i^*>},
\end{equation*}
hence, there is only one integer
\begin{equation*}
k_i=\bigg[c_i-\frac{<\sum\limits_{j=i+1}^n k_j \beta_j,\beta_i^*>}{<\beta_i^*,\beta_i^*>}\bigg]
\end{equation*}
such that $w_i\in [0,1)$, $\forall 1\leqslant i\leqslant n-1$. Above all, there is only one vector $w=\sum\limits_{i=1}^n c_i \beta_i^*-\sum\limits_{i=1}^n k_i \beta_i\in \mathcal{F}(I)$ such that $c\equiv w\ (\text{mod}\ I)$.
\end{proof}

Based on Lemma 2.6, we give an algorithm for the decryption of our general FHE scheme.\\

\noindent\rule[5pt]{16cm}{0.08em}

{\ \ \ \ \ \ \ \ \ \ \ \ \ \ \text{\bf{Algorithm 2: Decryption Algorithm for FHE Scheme}}}

\noindent\rule[5pt]{16cm}{0.08em}

$\bullet$\ \ \ For any ciphertext $c\in \mathbb{Z}^n$, given a full rank ideal lattice $I_i$ with HNF basis $B=$\\ \hspace*{1.2cm} $[\beta_1,\beta_2,\dots,\beta_n]$ and the corresponding orthogonal basis $B^*=[\beta_1^*,\beta_2^*,\dots,\beta_n^*]$, we\\ \hspace*{1.2cm} can find only one vector $\overline{u_i}=c-\sum\limits_{i=1}^n k_i \beta_i \in\mathcal{F}(I)$ such that $c\equiv \overline{u_i}\ (\text{mod}\ I_i)$, where
\begin{equation*}
k_n=\left[\frac{<c,\beta_n^*>}{<\beta_n^*,\beta_n^*>}\right],\ k_i=\bigg[\frac{<c,\beta_i^*>}{<\beta_i^*,\beta_i^*>}-\frac{<\sum\limits_{j=i+1}^n k_j \beta_j,\beta_i^*>}{<\beta_i^*,\beta_i^*>}\bigg],\ i=n-1,n-2,\dots,1.
\end{equation*}

\noindent\rule[5pt]{16cm}{0.05em}

\vspace{0.3cm}
To create an efficient algorithm for generating secret key of the above unbounded FHE scheme, we assume that $\phi(x)$ is an irreducible polynomial, so that the principal ideal $<\phi(x)>$ is a prime ideal in $\mathbb{Z}[x]$, and the quotient ring $\mathbb{Z}[x]/<\phi(x)>$ is a domain, equivalently, $(\mathbb{Z}^n,+,\otimes)$ becomes a domain. Under this assumption, we see that arbitrary many pairwise relatively prime ideals in $\mathbb{Z}^n$ almost come in automatically, we describe the process as follows.

\noindent\rule[5pt]{16cm}{0.08em}

{\ \ \ \ \ \ \ \ \ \ \ \ \ \ \text{\bf{Algorithm 3: Generating Algorithm for Secret Key}}}

\noindent\rule[5pt]{16cm}{0.08em}

$\bullet$\ \ \ \textbf{First step.} Randomly select an non-zero vector $\alpha\in \mathbb{Z}^n$ as input, and the output\\ \hspace*{3.4cm} is the following relatively prime two ideal $I_1$ and $I_2$, where
\begin{equation*}
I_1=<\alpha>,\ \text{and}\ I_2=<e-\alpha>.
\end{equation*}

$\bullet$\ \ \ \textbf{Second step.} Randomly select two non-zero vectors $\alpha_1\in I_1$ and $\alpha_2\in I_2$ as\\ \hspace*{3.8cm} input, and the output is the following ideal $I_3$, where
\begin{equation*}
I_3=<e-\alpha_1 \otimes \alpha_2>.
\end{equation*}
\hspace*{3.8cm} It is easy to see that $I_1,I_2,I_3$ are pairwise relatively prime ideals.

$\bullet$\ \ \ \textbf{Last step.} Suppose that $m-1$ pairwise relatively prime ideals $I_1,I_2,\dots,I_{m-1}$ are\\ \hspace*{3.3cm} selected, then one randomly finds $\alpha_1\in I_1$, $\alpha_2\in I_2$, $\dots$, $\alpha_{m-1}\in I_{m-1}$,\\ \hspace*{3.3cm} $\alpha_i\neq 0$, and the output ideal $I_m$ given by
\begin{equation*}
I_m=<e-\alpha_1 \otimes \alpha_2 \otimes \cdots \otimes \alpha_{m-1}>.
\end{equation*}
\hspace*{3.3cm} Obviously, $I_1,I_2,\dots,I_m$ are pairwise relatively prime ideals in $\mathbb{Z}^n$.

\noindent\rule[5pt]{16cm}{0.05em}

\vspace{0.3cm}
To construct an efficient algorithm for finding public key, we first show that

\textbf{Lemma 2.7}\ \ Suppose that $I_1,I_2,\dots,I_m$ are pairwise relatively ideals in $\mathbb{Z}^n$, and each $I_i$ is a principal ideal generated by $\alpha_i$, namely, $I_i=<\alpha_i>$. Let
\begin{equation*}
d_i=\mathop{\otimes}\limits_{j=1,j\neq i}^m \alpha_j,\ 1\leqslant i\leqslant m.
\end{equation*}
Then for every $d_i$, there is a vector $\mathcal{D}_i\in \mathbb{Z}^n$ such that
\begin{equation*}
d_i \otimes \mathcal{D}_i \equiv e\ (\text{mod}\ I_i),\ 1\leqslant i\leqslant m.
\end{equation*}

\begin{proof} Since $I_1,I_2,\dots,I_m$ are pairwise relatively prime by assumption, we have
\begin{equation*}
(I_i,I_1 I_2\cdots I_{i-1}I_{i+1}\cdots I_m)=1,\ 1\leqslant i\leqslant m.
\end{equation*}
In other words, we have $(I_i,<d_i>)=1$, or
\begin{equation*}
<\alpha_i>+<d_i>=\mathbb{Z}^n.
\end{equation*}
Therefore, there is a vector $\mathcal{D}_i\in \mathbb{Z}^n$ such that $d_i \otimes \mathcal{D}_i \equiv e\ (\text{mod}\ I_i)$, and we have Lemma 2.7 immediately.
\end{proof}

How to find the vector $\mathcal{D}_i$ defined by Lemma 2.7? We introduce a polynomial algorithm to obtain $\mathcal{D}_i$ ($1\leqslant i\leqslant m$), and generate the public key $\{A_1,A_2,\dots,A_m\}$ by taking $A_i=d_i\otimes \mathcal{D}_i$ ($1\leqslant i\leqslant m$).

\noindent\rule[5pt]{16cm}{0.08em}

{\ \ \ \ \ \ \ \ \ \ \ \ \ \ \text{\bf{Algorithm 4: Generating Algorithm for Public Key}}}

\noindent\rule[5pt]{16cm}{0.08em}

$\bullet$\ \ \ Let $I_i=<\alpha_i>$ $(1\leqslant i\leqslant m)$ be pairwise relatively prime, and
\begin{equation*}
d_i(x)=\mathop{\prod}\limits_{j=1,j\neq i}^m \tau^{-1}(\alpha_i),\ 1\leqslant i\leqslant m,
\end{equation*}
\hspace*{1.2cm} where the secret key $I_i=<\alpha_i>$ ($1\leqslant i\leqslant m$), and $\tau^{-1}$ is the inverse mapping of\\ \hspace*{1.2cm} $\tau$ given by (2.1).\\
\hspace*{1.2cm} Since $d_i(x)$ and $\tau^{-1} (\alpha_i)$ are relatively prime polynomial in quotient ring\\ \hspace*{1.2cm} $\mathbb{Z}[x]/<\phi(x)>$, it follows that
\begin{equation*}
<d_i(x)>+<\tau^{-1}(\alpha_i)>=\mathbb{Z}[x]/<\phi(x)>.
\end{equation*}
\hspace*{1.2cm} Therefore, there is a polynomial $\mathcal{D}_i(x)$ such that
\begin{equation*}
d_i(x)\mathcal{D}_i(x)\equiv 1\ (\text{mod}\ \tau^{-1}(\alpha_i)).
\end{equation*}
\hspace*{1.2cm} We put $\mathcal{D}_i=\tau(\mathcal{D}_i(x))$, and $A_i=d_i\otimes \mathcal{D}_i$, $1\leqslant i\leqslant m$.

\noindent\rule[5pt]{16cm}{0.05em}

\vspace{0.3cm}
To discuss the security of this scheme, we observe that all risks come from the encryption algorithm (2.6), we describe this risk as a generalized compact knapsack problem over the ring $(\mathbb{Z}^n,+,\otimes)$ as follows:
\begin{equation*}
\sum\limits_{i=1}^m a_i \otimes x_i=u,\quad |x_i|\leqslant \beta, \tag{2.7}
\end{equation*}
where $a_1,a_2,\cdots,a_m$ are given $m$ vectors in $\mathbb{Z}^n$, $u\in \mathbb{Z}^n$ is a target vector, and $\beta=\max\limits_{1\leqslant i\leqslant m} t_i$. We will show that the security next section.

On the other hands, for any $a\in \mathbb{Z}$, we see that $H^*(\overline{a})=aI_n$. Therefore, this problem may transfer to the standard Ring-SIS problem: suppose that $A=[A_1,A_2,\cdots,A_m]\in \mathbb{Z}^{n\times m}$, where $A_i$ is the public key given by (2.5), $u\in \mathbb{Z}^n$ is a target vector, find $x=\begin{pmatrix} x_1\\ \vdots\\ x_n \end{pmatrix}\in \mathbb{Z}^n$ such that $Ax=u$, and $0<|x|\leqslant \beta$. Let $q$ be a sufficiently large positive integer, this problem can be changed to an inhomogeneous version of the SIS problem, which is to find a short integer solution to $Ax\equiv u\ (\text{mod}\ q)$. It is not hard to show that the homogeneous and inhomogeneous problem are essentially equivalent for typical parameters.

According to Ajtai's seminal work \cite{02,03}, the hardness of the SIS problem relative to the worst-case lattice problem. More precisely, for any $m=\text{poly}(n)$, any $\beta>0$, and any sufficiently large $q\geqslant \beta\cdot \text{poly}(n)$, solving $\text{SIS}_{n,q,\beta,m}$ with non-negligible probability is at least as hard as solving the decisional approximate shortest vector problem $\text{GapSVP}_{\gamma}$ for some $\gamma=\beta\cdot\text{poly}(n)$ (also see \cite{25} and Theorem 4.1.2 of \cite{28}). Therefore, the secure of our FHE scheme is solely under a worst-case complexity assumption, this is a reasonable solution to {\bf{Open Question 11}} of \cite{28}.

\section{The Generalized Compact Knapsack Problem Over $\mathbb{Z}^n$}

In this section, we discuss the secure of our scheme based on the general compact knapsack problem over the ring $(\mathbb{Z}^n,+,\otimes)$. In \cite{22}, Micciancio has proved that if we can solve the knapsack problem over $\mathbb{Z}_q^n$ for some sufficiently large positive integer number $q$, then there is a probabilistic polynomial algorithm solving the covering radius problem for any $n$ dimensional full rank cyclic lattice. First we generalize Micciancio's result to arbitrary ideal lattices based on our precious work \cite{39}, and then solving the knapsack problem from $\mathbb{Z}_q^n$ to $(\mathbb{Z}^n,+,\otimes)$. We give an entire proof for the reason of completeness, although the method we present here is quite similar to Micciancio's original proof.

\vspace{0.3cm}
\textbf{DEFINITION 3.1} \ \ Let $L$ be a full rank lattice, $\gamma(n)$ is a parameter of $n$, $\gamma(n)\geqslant 1$, $\text{CDP}_{\gamma}$ problem is to find an $r$ such that
\begin{equation*}
\rho(L)\leqslant r\leqslant \gamma(n)\rho(L),
\end{equation*}
where $\rho(L)=\max\limits_{x\in \mathbb{R}^n} \text{dist}(x,L)$ and $\text{dist}(x,L)=\min\limits_{\alpha\in L}|x-\alpha|$.

\vspace{0.3cm}

\textbf{DEFINITION 3.2} \ \ Let $L$ be a full rank lattice, $S=\{s_1,s_2,\cdots,s_n\}\subset L$ be $n$ linearly independent lattice vectors. $S^*=\{s_1^*,s_2^*,\cdots,s_n^*\}$ is the orthogonal basis corresponding to $S$ by the Gram-Schmidt method. We define
\begin{equation*}
\sigma(S)=\Big(\sum\limits_{i=1}^{n} |s_i^*|^2\Big)^{\frac{1}{2}}.
\end{equation*}

Here we give our main result to show that the generalized knapsack problem over $\mathbb{Z}^n$ is at least as hard as the covering radius problem for any $n$ dimensional full rank ideal lattice.

\textbf{THEOREM 2} \ \ Let $m=O(\log n)$, $k=\tilde{O}(\text{log}n)$, $\phi_{\max}=\max\{|\phi_0|,|\phi_1|,\dots,|\phi_{n-1}|\}$, $M=\sqrt{2+2\phi_{\max}^2}$, $W=\frac{M^n-1}{M-1}$, $\gamma\geqslant 16mkn^3 W$, if we can solve the knapsack problem (2.7), then there is a positive probabilistic polynomial algorithm solving the covering radius problem $\text{CDP}_{\gamma}$ for any $n$ dimensional full rank ideal lattice $L$.

\vspace{0.3cm}
Remark: In the original work of Micciancio \cite{22}, the parameter $\gamma$ is bigger than $16mkn^3$, we require $\gamma\geqslant 16mkn^3 W$, where $W$ is given by $\frac{M^n-1}{M-1}$. The main difference is to estimate the length of convolutional product $\alpha\otimes\beta$ for any two vectors $\alpha$ and $\beta$. It is clearly that $|\alpha\otimes\beta|\leqslant \sqrt{n}|\alpha||\beta|$ in the case of circulant lattice, but it is non-trivial in the case of ideal lattices.

\vspace{0.3cm}
\textbf{Lemma 3.1}\ \ For any $\alpha,\beta\in \mathbb{Z}^n$, we have
\begin{equation*}
|\alpha\otimes\beta|\leqslant W|\alpha|\cdot|\beta|,
\end{equation*}
where $W$ is defined in Theorem 2.

\begin{proof}

We first prove $|H\alpha|\leqslant M|\alpha|$. Let $\alpha=(\alpha_0,\alpha_1,\dots,\alpha_{n-1})^{T}$, then
\begin{align*}
|H\alpha|^2 &=\phi_0^2 \alpha_0^2+(\alpha_0+\phi_1 \alpha_1)^2+\cdots+(\alpha_{n-2}+\phi_{n-1}\alpha_{n-1})^2 \\
&\leqslant \phi_0^2 \alpha_0^2+2(\alpha_0^2+\phi_1^2 \alpha_1^2)+\cdots+2(\alpha_{n-2}^2+\phi_{n-1}^2 \alpha_{n-1}^2) \\
&\leqslant 2(\alpha_0^2+\cdots+\alpha_{n-2}^2)+2\phi_{\max}^2(\alpha_0^2+\alpha_1^2+\cdots+\alpha_{n-1}^2) \\
&\leqslant (2+2\phi_{\max}^2)|\alpha|^2.
\end{align*}
So $|H\alpha|\leqslant M|\alpha|$. Similarly,
\begin{equation*}
|H^2 \alpha|\leqslant M |H\alpha|\leqslant M^2 |\alpha|.
\end{equation*}
In the same way, we can get $|H^k \alpha|\leqslant M^k |\alpha|$, $\forall 1\leqslant k\leqslant n-1$. Let $\beta=(\beta_0,\beta_1,\dots,\beta_{n-1})^{T}$, it follows that
\begin{align*}
|\alpha\otimes\beta| &=|H^*(\alpha)\beta|=|\beta_0 \alpha+\beta_1 H\alpha+\cdots+\beta_{n-1} H^{n-1}\alpha| \\
& \leqslant \sum\limits_{i=0}^{n-1} |\beta_{i} H^i \alpha|\leqslant \sum\limits_{i=0}^{n-1} M^i |\alpha| |\beta|=\frac{M^n-1}{M-1}|\alpha|\cdot|\beta|.
\end{align*}
We complete this proof.
\end{proof}

Let $L=\mathcal{L}(B)\subset \mathbb{R}^n$ be a full rank ideal lattice, $q\geqslant 4mkn^2 W^2$, $\{e_0,e_1,\cdots,e_{n-1}\}\subset \mathbb{Z}_q^n$ is a standard orthogonal basis, $S=\{s_1,s_2,\cdots,s_n\}\subset L$ is a set of $n$ linearly independent vectors. We define the parameter
\begin{equation*}
\mu=\left(\frac{4nq}{\gamma}+\frac{W}{2}\right)\sigma(S). \tag{3.1}
\end{equation*}
According to Lemma 1.6 in Chapter 3 in \cite{36}, there is a lattice vector $c\in L$ such that $|c-\mu e_0|\leqslant \frac{1}{2}\sigma(S)$, let $B'=q(H^*(c))^{-1}B$. It follows that the lattice $\mathcal{L}(B')$ generated by $B'$ satisfies $q\mathbb{Z}^n\subset \mathcal{L}(B')$ according to Lemma 2.1 in Chapter 3 in \cite{36}. Therefore, $q\mathbb{Z}^n$ is an additive subgroup in $\mathcal{L}(B')$. Randomly choose $mk$ elements $x_{ij}^{'}\ (1\leqslant i\leqslant m,1\leqslant j\leqslant k)$ in the quotient group $G=\mathcal{L}(B')/q\mathbb{Z}^n$, the integral vectors $w_{ij}^{'}$ of $x_{ij}^{'}$ is defined by
\begin{equation*}
w_{ij}^{'}=\lfloor x_{ij}^{'}\rceil\in \mathbb{Z}^n,\ 1\leqslant i\leqslant m,\ 1\leqslant j\leqslant k.
\end{equation*}
Let
\begin{equation*}
a_{ij}\equiv w_{ij}^{'}\ (\text{mod}\ q),\ \  a_i\equiv \sum\limits_{j=1}^{k} w_{ij}^{'}\ (\text{mod}\ q),
\end{equation*}
Assume $A=(a_1,a_2,\cdots,a_m)\in \mathbb{Z}_{q}^{n\times m}$. Since the knapsack problem (2.7) is solvable on $\mathbb{Z}^n$, it could also be solved on $\mathbb{Z}_q^n$. Let $y=(y_1,y_2,\cdots,y_m)\in \mathbb{Z}_{q}^{n\times m}$ and $\hat{y}=(\hat{y}_1,\hat{y}_2,\cdots,\hat{y}_m)\in \mathbb{Z}_{q}^{n\times m}$ be two different integer matrices such that $\sum\limits_{i=1}^m a_i\otimes (y_i-\hat{y}_i)=0,\ |y_i|\leqslant \sqrt{n},\ |\hat{y}_i|\leqslant \sqrt{n}, \forall 1\leqslant i\leqslant m$. We define
\begin{equation*}
x_{ij}=\frac{1}{q}H^*(c)x_{ij}^{'}, w_{ij}=\frac{1}{q}H^*(c)w_{ij}^{'}, 1\leqslant i\leqslant m, 1\leqslant j\leqslant k,
\end{equation*}
and
\begin{equation*}
s'=\sum\limits_{i=1}^{m} \sum\limits_{j=1}^{k} (x_{ij}-w_{ij}) \otimes (y_{i}-\hat{y}_{i}). \tag{3.2}
\end{equation*}
Then $x_{ij}$ is a lattice vector in the given ideal lattice $L=\mathcal{L}(B)$ ($1\leqslant i\leqslant m,1\leqslant j\leqslant k$), and $s'$ is also a lattice vector in $L=\mathcal{L}(B)$ based on Lemma 2.2 in Chapter 3 in \cite{36}.

The next lemma gives an estimation of the length of $s'$, which has some differences from the proof of Micciancio's.

\vspace{0.2cm}
\textbf{Lemma 3.2} \ \ Let $S=\left\{s_1,s_2,\cdots,s_n\right\}\subset L$ be a set of $n$ linearly independent vectors in the full rank ideal lattice $L$. Denote $|S|=\max\limits_{1\leqslant i\leqslant n} |s_i|$, $s'$ is the lattice vector defined in (3.2), then
\begin{equation*}
|s'|\leqslant\frac{1}{2}|S|.
\end{equation*}

\begin{proof} This proof is similar to that of Lemma 2.3 in Chapter 3 in \cite{36} except some computations about the parameters. We prove $|s'|\leqslant \frac{1}{2\sqrt{n}}\sigma(S)$ first. Based on the definition of $s'$ in (3.2),
\begin{equation*}
|s'|\leqslant \sum\limits_{i=1}^m \sum\limits_{j=1}^k \big|(x_{ij}-w_{ij})\otimes (y_i-\hat{y}_i)\big|. \tag{3.3}
\end{equation*}
\begin{equation*}
x_{ij}-w_{ij}=\frac{1}{q}H^*(c)(x_{ij}^{'}-w_{ij}^{'})=\frac{1}{q}c \otimes (x_{ij}^{'}-w_{ij}^{'}).
\end{equation*}
Let $\alpha=c-\mu e_0$, where $\mu$ is defined in (3.1). Then $|\alpha|\leqslant \frac{1}{2}\sigma(S)$, and
\begin{equation*}
x_{ij}-w_{ij}=\frac{1}{q}(\alpha+\mu e_0)\otimes (x_{ij}^{'}-w_{ij}^{'})=\frac{1}{q}H^*(\alpha+\mu e_0)(x_{ij}^{'}-w_{ij}^{'})
\end{equation*}
\begin{equation*}
\ \ =\frac{1}{q}\mu H^*(e_0) (x_{ij}^{'}-w_{ij}^{'})+\frac{1}{q} H^*(\alpha) (x_{ij}^{'}-w_{ij}^{'})\quad
\end{equation*}
\begin{equation*}
\ =\frac{\mu}{q}(x_{ij}^{'}-w_{ij}^{'})+\frac{1}{q} H^*(\alpha) (x_{ij}^{'}-w_{ij}^{'}).\qquad\qquad
\end{equation*}
Since $w_{ij}^{'}=\lfloor x_{ij}^{'}\rceil\in \mathbb{Z}^n$, we have
\begin{equation*}
|x_{ij}^{'}-w_{ij}^{'}|\leqslant \frac{1}{2}\sqrt{n},
\end{equation*}
combine with Lemma 3.1, we get
\begin{equation*}
|x_{ij}-w_{ij}|\leqslant \frac{\mu}{q}|x_{ij}^{'}-w_{ij}^{'}|+\frac{1}{q} |\alpha\otimes (x_{ij}^{'}-w_{ij}^{'})|
\end{equation*}
\begin{equation*}
\qquad\qquad\leqslant \frac{\mu}{q}\cdot\frac{1}{2}\sqrt{n}+\frac{1}{q}\cdot W\cdot\frac{1}{2}\sigma(S)\cdot\frac{\sqrt{n}}{2}
\end{equation*}
\begin{equation*}
\qquad\qquad\qquad\ \ =\frac{\sqrt{n}}{2q}\left(\frac{4nq}{\gamma}+\frac{W}{2}\right)\sigma(S)+\frac{\sqrt{n}W}{4q}\sigma(S)
\end{equation*}
\begin{equation*}
\qquad\ \ =\sigma(S)\left(\frac{2n^{\frac{3}{2}}}{\gamma}+\frac{\sqrt{n}W}{2q}\right)\qquad\
\end{equation*}
\begin{equation*}
\qquad\qquad\qquad\ \leqslant \sigma(S)\left(\frac{1}{8}\cdot\frac{1}{mkn^{\frac{3}{2}}W}+\frac{1}{8}\cdot\frac{1}{mkn^{\frac{3}{2}}W}\right)
\end{equation*}
\begin{equation*}
\ \ \ =\sigma(S)\frac{1}{4mkn^{\frac{3}{2}}W}.\qquad\quad\ \
\end{equation*}
Based on (3.3), we know
\begin{equation*}
|s'|\leqslant mkW\max\limits_{i,j}|x_{ij}-w_{ij}|\cdot \max\limits_{i}|y_i-\hat{y}_i|
\end{equation*}
\begin{equation*}
\ \ \leqslant mkW\cdot \frac{\sigma(S)}{4mkn^{\frac{3}{2}}W}\cdot 2\sqrt{n}=\frac{\sigma(S)}{2\sqrt{n}}.\
\end{equation*}
Since
\begin{equation*}
\sigma(S)\leqslant \left(\sum\limits_{i=1}^n |s_i|^2\right)^{\frac{1}{2}}\leqslant \sqrt{n} |S|,
\end{equation*}
We can get
\begin{equation*}
|s'|\leqslant \frac{\sigma(S)}{2\sqrt{n}}\leqslant \frac{\sqrt{n}|S|}{2\sqrt{n}}=\frac{1}{2}|S|.
\end{equation*}
So we complete the proof of Lemma 3.2.

\end{proof}

Based on the above lemmas, Theorem 2 follows directly from Micciancio's method \cite{22}. From Theorem 2, we can see that the general compact knapsack problem over $\mathbb{Z}^n$ is at least as hard as the covering radius problem $\text{CDP}_{\gamma}$ of any ideal lattice. Therefore, the security of our unbounded FHE scheme is solely under the worst-case complexity assumption, as a result, it is a reasonable solution to {\bf {Open Question 11}} of \cite{28}.

\section{A Practical System}

As an example, we introduce a practical system for FHE using some basic results about the cyclotomic field \cite{34}. Suppose that $p$ is an odd prime number and $Q\left(\xi_p\right)$ is the cyclotomic field, where $\xi_p=e^{\frac{2 \pi i}{p}}$ is a primitive $p$-th root of unit. Let
\begin{equation*}
\mathbb{Z}\left[\xi_p\right]=\left\{\sum_{i=0}^{p-1} a_i \xi_p^i \mid a_i \in \mathbb{Z}\right\}.
\end{equation*}
It is known that $\mathbb{Z}\left[\xi_p\right]$ is the ring of algebraic integers of field $Q(\xi_{ p})$. Therefore $\mathbb{Z}\left[\xi_p\right]$  is a Dedekind domain (so we have unique factorization into prime ideals, etc. see Proposition $1.2$ of \cite{34}).

To construct an commutative ring for $\mathbb{Z}^n$, we select $\phi(x)= x^{p-1}+x^{p-2}+\cdots+x+1 \in \mathbb{Z}[x]$. Obviously, $\phi(x)$ is the minimal polynomial of $\xi_{p} . $ Let $ n=p-1$, then we obtain a ring $\left(\mathbb{Z}^n,+,\otimes\right)$ in terms of  $\phi(x)$ and the rotation matrix $H_\phi$, it is easy to see that (see (3.2) of \cite{37})
$$
\left(\mathbb{Z}^n,+,\otimes\right)\cong  \mathbb{Z}[x] /\langle\phi(x)\rangle \cong \mathbb{Z}\left[\xi_p\right].
$$
Thus, $\mathbb{Z}^n$ becomes a Dedekind domain.

For any integer $q\in \mathbb{Z}$, we define an integer vector $\alpha_q\in \mathbb{Z}^n$ by
\begin{equation*}
\alpha_q=\left(\begin{array}{l}
q \\
0 \\
\vdots\\
0 \\
1
\end{array}\right) \in \mathbb{Z}^n.
\end{equation*}

The principal ideal $\left\langle\alpha_q\right\rangle $ generated by $\alpha_q$ is denoted by
\begin{equation*}
I_q=\left\langle\alpha_q\right\rangle=\mathcal{L}\left(H^*(\alpha_q)\right) .
\end{equation*}

\textbf{Lemma 4.1}\ \
 If $q_1$ and $q_2$ are two different prime numbers, then $I_{q_1}$ and $I_{q_2}$ are relatively prime ideal lattices in $\mathbb{Z}^n$.

  \begin{proof}
  Suppose that $q$ is a prime number. Since $\left(\mathbb{Z}^n,+,\otimes\right)\cong  \mathbb{Z}[x] /\langle\phi(x)\rangle,$ we see  the polynomial $\tau^{-1}\left(\alpha_q\right)=\alpha_q(x)=x^{n-1}+q\in \mathbb{Z}[x]$, which is the type polynomial of Eisenstein, thus it is an irreducible polynomial over $\mathbb{Z}[x]$. Regarding $\alpha_{q}(x)$ as the polynomial of $\mathbb{Z}[x] /\langle\phi(x)\rangle$, clearly, it is also an irreducible polynomial in $\mathbb{Z}[x] /\langle\phi(x)\rangle$. By  assumption, $q_{1}$ and $q_2$ are different prime numbers, we show that the principal ideals $\left\langle\alpha_{q_1}(x)\right\rangle$, and $\left\langle\alpha_{q_2}(x)\right\rangle$ are relatively prime ideals in $\mathbb{Z}[x] /\langle\phi(x)\rangle$. Since $\mathbb{Z}[x] /\langle\phi(x)\rangle$ is a Dedekind domain, if
  $\left\langle\alpha_{q_1}(x)\right\rangle$ and $\left\langle\alpha_{q_2}(x)\right\rangle$  are not relatively prime, then, there exists a prime ideal $P$ in $\mathbb{Z}[x] /\langle\phi(x)\rangle$ such that
$$
\left\langle\alpha_{q_{1}}(x)\right\rangle \subset P,\  \text { and } \ \left\langle\alpha_{q_{2}}(x)\right\rangle \subset P.
$$
It follows for any positive integer $k \geq 1$ that
$$
\left<\alpha^{k}_{q_{1}}(x)\right> \subset P^{k},\  \text { and } \ \left\langle\alpha^{k}_{q_{2}}(x)\right\rangle \subset P^{k}.
$$
It is known that three exists a positive integer  $k \geqslant 1$, such that $P^k=\langle d ( x)\rangle$ is a principal ideal, we thus have
$$
\left\langle\alpha^{k}_{q_{1}}(x)\right\rangle \subset \langle d ( x)\rangle,\  \text { and } \ \left\langle\alpha^{k}_{q_{2}}(x)\right\rangle \subset \langle d ( x)\rangle.
$$
In other words, we have $d(x) \mid \alpha^{k}_{q_{1}}(x)$, and $d(x) \mid \alpha^{k}_{q_{2}}(x)$. However, this is impossible, since $\alpha_{q_1}(x)$ and $\alpha_{q_2}(x)$ are two irreducible polynomials in $\mathbb{Z}[x] /\langle\phi(x)\rangle$, and $\alpha_{q_{1}}(x) \neq \alpha_{q_{2}}(x)$.

This proves that $\left\langle\alpha_{q_{1}}(x)\right\rangle$ and $\left\langle\alpha_{q_{2}}(x)\right\rangle$ are relatively prime  ideals in
$\mathbb{Z}[x] /\langle\phi(x)\rangle$. Equivalently, $I_{q_{1}}$ and $I_{q_{2}}$ are two relatively prime ideal lattices in $\mathbb{Z}^{n}.$
\end{proof}

\textbf{Lemma 4.2}\ \   The one dimensional modulus of the principal ideal $I_q$ is
\begin{equation*}
t_q=t(I_{q})= q^{n}-q^{n-1}+\cdots-q+1 .
\end{equation*}

\begin{proof}
It is not hard to compute that $\text{det}(H^{*}(\alpha_{q}))=q^{n}-q^{n-1}+\cdots-q+1.$ Since the polynomial $ x^{n}-x^{n-1}+\cdots-x+1$ is an irreducible polynomial over $\mathbb{Z}$, we have the assertion of lemma immediately.
\end{proof}

According to Lemma 4.1 and Lemma 4.2,  we obtain a generating algorithm for the secret key as follows:

\begin{table}[h]
\begin{tabular}{l}
\hline
  {{\textbf{ \ \ \ \ \ \ \ \ \ \ \ \ \ \ \ \ \ \ \ Algorithm 5: Generating Algorithm for Secret Key}}}\\
\hline
\\
$\bullet$\  \ Randomly select $m$ different prime numbers $q_{1}$, $ q_{2}$, $\cdots,$ $q_{m}$ such that $q_{i}\neq p$. The
principle  \\
\ \ \ \ ideal lattices \\
\ \ \ \ \ \ \ \ \ \ \ \ \ \ \ \ \ \ \ \ \ \ \ \ \ \ \ \ \ \ \ \ \ \ \ \ \ \ \ \ \ \ \ \
$
I_{q_{i}}=\langle \alpha_{q_{i}}\rangle,\ \ 1\leq i \leq m
$ \\
\ \ \ \ are the secret key, and the corresponding one dimensional modulus $t_{q_{i}}$ given by\\
 \ \ \ \ \ \ \ \ \ \ \ \ \ \ \ \ \ \ \ \ \ \ \ \ \ \ \ \ \ \ \ \ \ \
 $
t_{q_{i}}=t(I_{q_{i}} )= q_{i}^{n}-q_{i}^{n-1}+\cdots-q_{i}+1. $\ \ \ \ \ \ \ \ \ \ \ \ \ \ \ \ \ \ \ \ \ \ \ \ \ \ \ \ \ \ \ \ \ (4.1) \\ \\
\hline
\end{tabular}
\end{table}

\newpage
To get an attainable algorithm for public key, we  define $m$ vectors $d_i(1\leq i \leq m)$ by

\begin{equation*}
 d_i=\alpha_{q_1} \otimes \alpha_{q_2} \otimes \cdots \otimes \alpha_{q_{i-1}} \otimes \alpha_{q_{i+1}} \otimes \cdots \otimes \alpha_{q_{m}} \\
 =\otimes_{\substack{j=1 \\
j \neq i}}^m \alpha_{q_{j}}, \quad 1 \leq i \leq m.
\end{equation*}

\vspace{0.3cm}

\textbf{Lemma 4.3}\ \   For every vector $d_i(1 \leq i \leq m)$, there is a vector $\mathcal{D}_i \in \mathbb{Z}^n$ such that
\begin{equation*}
d_i\otimes \mathcal{D}_i \equiv e\left(\text{mod}\  I_{q_i}\right),
\end{equation*}
where $e=\left(\begin{array}{l}1 \\ 0 \\ \vdots \\ 0\end{array}\right)$ is the unit element  of rimg $\left(\mathbb{Z}^{n},+,\otimes\right)$.

\begin{proof}
By lemma 4.1, $I_{q_1}, I_{q_2}, \cdots, I_{q_m}$ are pairwise relatively
prime ideals, hence we have \\
$\left(I_{q_i}, I_{q_1} I_{q_2} \cdot I_{q_{i-1}} I_{q_{i+1}} \cdot I_{q_m}\right)=1$.
In other words, we have $\left(\left\langle\alpha_{q_{i}}\right\rangle,\left\langle d_i\right\rangle\right)=1$, or
$$
 \left\langle\alpha_{q_i}\right\rangle+\left\langle d_{i} \right\rangle=\mathbb{Z}^{n}.
$$
Therefore, there is a vector $\mathcal{D}_i \in \mathbb{Z}^{n}$ such that $\mathcal{D}_i \otimes  d_i \equiv e (\text{mod}\  I_{q_{i}} )$. we have Lemma $4.3$.

\end{proof}

We propose a polynomial algorithm to find the vector $\mathcal{D}_i$ appeared in Lemma 4.3, and then put $A_i=d_i\otimes \mathcal{D}_i$ getting the public key $A_1, A_2, \cdots, A_m$. The polynomial algorithm for finding vector $\mathcal{D}_i$ like follows.

\vspace{0.5cm}
\begin{table}[h]
\begin{tabular}{l}
\hline
  {{\textbf{\ \ \  \ \ \ \ \ \ \ \ \ \ \ \ \ \ \ \ Algorithm 6: Generating Algorithm for Public Key }}}\\
\hline
\\
$\bullet$\ \ Let \\
\ \ \ \ \ \ \ \ \ \ \ \ \ \ \ \ \ \ \ \ \ \ \ \ \ \ \ \ \ \ \ \
$
d_i(x)=\prod_{\substack{j=1 \\j \neq i}}^m\left(x^{n-1}+q_j\right), 1 \leq i \leq m.
$ \\
\ \ \ \ Since polynomial $x^{n-1}+q_i$ and $x^{n-1}+q_j(j \neq i)$ are relatively prime in $\mathbb{Z}[x] /\langle\phi(x)\rangle$, \\
\ \ \ \ it follows that $\left(x^{n-1}+q_i, d_i(x)\right)=1$. Therefore, there is a polynomial $\mathcal{D}_{i}(x)$ such that\\
\ \ \ \ \ \ \ \ \ \ \ \ \ \ \ \ \ \ \ \ \ \ \ \ \ \ \ \ \ \ \ \ \ \ \
$
d_i(x) \mathcal{D}_i(x) \equiv 1\left(\text{mod}\  (x^{n-1}+q_i )\right).
$\\
\ \ \ \ We obtain $\mathcal{D}_i=\tau\left(\mathcal{D}_i(x)\right)$ under the mapping $\tau$ given in section 2. \\ \\
\hline
\end{tabular}
\end{table}

Now, we describe an attainable algorithm for our  FHE Scheme based on cyclotomic field as follows.

\vspace{0.3cm}
\begin{table}[h]
\begin{tabular}{l}
\hline
  {{\textbf{ \ \ \ \ \ \ Algorithm 7: The Attainable Algorithm for Unbounded FHE Scheme}}}\\
\hline
\\
$\bullet$ \ \textbf{Secret key: } Let $q_1, q_2, \cdots, q_m$ be $m$ different prime numbers such that $q_i \neq p$. The $m$  \\
\ \ \ \ \ \ \ \ \ \ \ \ \ \ \ \ \ \ \ \ \ \   pairwise relatively prime principal  ideal lattices $I_{q_1}, I_{q_2}, \cdots, I_{q_m}$ are\\
\ \ \ \ \ \ \ \ \ \ \ \ \ \ \ \ \ \ \ \ \ \    the secret keys for decryption.\\

$\bullet$\ \textbf{Public key: }  Let $A_i=d_i \otimes \mathcal{D}_i\left(1 \leqslant i \leqslant m\right)$, then $A_1, A_2, \cdots, A_m$ are the public key for  \\
\ \ \ \ \ \ \ \ \ \ \ \ \ \ \ \ \ \ \ \ \ \  encryption.  The plaintext space is \\
\ \ \ \ \ \ \ \ \ \ \ \ \ \ \ \ \ \ \ \ \ \ \ \ \ \ \ \ \ \ \ \ \ \ \  \ \ \ \ \
$
\mathcal{P}=\mathbb{Z}_{t_1}\bigoplus \mathbb{Z}_{t_2}\bigoplus\cdots\bigoplus \mathbb{Z}_{t_m},
$\\
\ \ \ \ \ \ \ \ \ \ \ \ \ \ \ \ \ \ \ \ \ \ where $t_i$ is the one dimensional modulus of $I_{q_i}$ given by (4.1).\\

$\bullet$\ \textbf{Encryption: } For any plaintext $u=\left(u_1, u_2, \cdots , u_m\right) \in \mathop{\bigoplus}\limits_{i=1}^m\mathbb{Z}_{t_{i}}$, then\\
\ \ \ \ \ \ \ \ \ \ \ \ \ \ \ \ \ \ \ \ \ \ \ \ \ \ \ \ \ \ \ \ \ \ \  \ \ \ \ \
 $
c=f(u)=\bar{u}_1 \otimes A_1+\bar{u}_2 \otimes A_2+\cdots+\bar{u}_m\otimes A_m,
$\\
\ \ \ \ \ \ \ \ \ \ \ \ \ \ \ \ \ \ \ \ \ \
where $\bar{u}_i$ is the embedding of $u_i$ into $\mathbb{Z}^n$.\\

$\bullet$\ \textbf{Decryption: }  For given ciphertext $c \in \mathbb{Z}^n$, there  is an unique vector $\bar{u}_i$ in the orthogonal \\
\ \ \ \ \ \ \ \ \ \ \ \ \ \ \ \ \ \ \ \ \ \
parallelepiped $\mathcal{F}\left(I_{q_i}\right)$, such that $c\equiv\bar{u}_i (\text{mod}\ I_{q_i})$, $1\leq i \leq m$. Thus, one has \\
\ \ \ \ \ \ \ \ \ \ \ \ \ \ \ \ \ \ \ \ \ \ \ \ \ \ \ \ \ \ \ \ \ \ \  \ \ \ \ \
$
f^{-1}(c)=\left(u_1, u_2, \cdots, u_m\right)=u.
$ \\ \\
\hline
\end{tabular}
\end{table}

The property of FHE follows immediately from the general construction. We mainly discuss the security of this practical system for FHE. Obviously, an additional  risk comes from the messages of one dimensional modulus $t_i\left(1 \leq i \leq m\right)$. It is equivalent  to find a solution of the algebraic equation with high degree of
$$
x^n-x^{n-1}+\cdots-x+1=t
$$
for given target value $t\in\mathbb{Z}$. This is an ancient hard problem in mathematics, there is  no general method to find the solution according to the famous Galois theory. Therefore, we conclude that there are no special risks coming from the one dimensional modulus $t_i$ of $I_{q_i}$.

\section{Conclusions}

In this work, we construct the first unbounded fully homomorphic encryption scheme without using bootstrapping transformation technique. It remains a problem to construct a probabilistic algorithm for generating secret key and public key, so that the security of our scheme solely depends upon short integer solution problem (SIS), or R-SIS. To our best knowledge, there is no a cryptosystem directly connect with SIS or R-SIS. On the other hands, the fully homomorphic signature is the dual question of the fully homomorphic encryption, we will develop a scheme for fully homomorphic signature in the following works. \\ \\

\textbf{Acknowledgements}\ \ This work was prepared during our visiting to Henan Academy of Sciences, the authors appreciate professor Tianze Wang and professor Jingluo Huang for invitation and hosting.

\end{document}